\documentclass[%
 reprint,
 amsmath,amssymb,
 aps, prd,
showkeys
]{revtex4-2}

\usepackage{graphicx}
\usepackage{dcolumn}
\usepackage{bm}

\usepackage{color}
\usepackage{float}
\usepackage[normalem]{ulem}

\usepackage{multirow}
\usepackage{lineno}
\usepackage{calc}

\begin{document}

\title{Direct determination of the $^{235}$U to $^{239}$Pu inverse beta decay yield ratio in the power reactor neutrino experiments.}

\author{I.~Alekseev$^{a,b,c,d}$}
\author{V.~Belov$^{d,e}$}
\author{A.~Bystryakov$^{b,e,f}$}
\author{M.~Danilov$^{b,d}$}
\author{D.~Filosofov$^{e}$}
\author{M.~Fomina$^{e}$}
\author{P.~Gorovtsov$^{c}$}
\author{Ye.~Iusko$^{a,g}$}
\author{S.~Kazartsev$^{d,e}$}
\author{V.~Khvatov$^h$}
\author{S.~Kiselev$^h$}
\author{A.~Kobyakin$^{a,b,c}$}
\author{A.~Krapiva$^{a,b,c}$}
\author{A.~Kuznetsov$^{e}$}
\author{I.~Machikhiliyan$^{i}$}
\author{N.~Mashin$^{a,b,c}$}
\author{D.~Medvedev$^{e}$}
\author{V.~Nesterov$^{a,b}$}
\author{D.~Ponomarev$^{b,d,e}$}
\author{I.~Rozova$^{e}$}
\author{N.~Rumyantseva$^{a,e,f}$}
\author{V.~Rusinov$^{a,b}$}
\author{E.~Samigullin$^{a,b}$}
\author{Ye.~Shevchik$^{e}$}
\author{M.~Shirchenko$^{d,e}$}
\author{Yu.~Shitov$^{j}$}
\author{N.~Skrobova$^{a,b,d}$}
\author{D.~Svirida$^{a,b,d}$}
\author{E.~Tarkovsky$^{a}$}
\author{E.~Yakushev$^{e}$}
\author{I.~Zhitnikov$^{d,e}$}
\author{A.~Yakovleva$^{c}$}
\author{D.~Zinatulina$^{e,k}$}
 
\affiliation{$^a$National Research Center "Kurchatov Institute",\\ Akademik Kurchatov square 1, Moscow, 123182, Russia}
\affiliation{$^b$Lebedev Physical Institute of the Russian Academy of Sciences, \\ Leninskiy avenue 53, Moscow, 119991, Russia}
\affiliation{$^c$Moscow Institute of Physics and Technology, \\ Institutskiy lane 9, Dolgoprudny, Moscow Region, 141701, Russia}
\affiliation{$^d$Institute for Nuclear Research of the Russian Academy of Sciences,\\ 60th October Anniversary Prospect 7a, Moscow 117312, Russia}
\affiliation{$^e$Joint Institute for Nuclear Research, \\ Joliot-Curie str. 6, Dubna, Moscow region, 141980, Russia}
\affiliation{$^f$Dubna State University, Universitetskaya str. 19, Dubna, Moscow region, 141982, Russia}
\affiliation{$^g$National Research Nuclear University Moscow Engineering Physics Institute, \\ Kashirskoe shosse 31, Moscow, 115409, Russia}
\affiliation{$^h$JSC Rosenergoatom Concern Affiliate—Kalinin Nuclear Power Plant, Udomlya, 171841 Russia}
\affiliation{$^i$Federal State Unitary Enterprise Dukhov Automatics Research Institute, \\ Sushchevskaya str. 22, Moscow, 127055, Russia}
\affiliation{$^j$Institute of Experimental and Applied Physics, Czech Technical University in Prague, \\ Husova 240/5, Prague, 110 00 Czech Republic} 
\affiliation{$^k$Voronezh State University, \\ Universitetskaya square 1, Voronezh, 1394018, Russia}

\date{\today}

\def\antinu{$\tilde{\nu_e}$}

\begin{abstract}
The yields of the inverse beta decay events produced by antineutrinos from a certain nuclear reactor fuel component are used by many experiments to check various model
predictions. Yet measurements of the absolute yields feature significant uncertainties coming, mainly,  from the understanding of the antineutrino detection efficiency.
This work presents a simple novel approach to directly determine the $^{235}$U to $^{239}$Pu inverse beta decay yield ratio using the fuel evolution analysis. This 
ratio can be used for a sensitive test of reactor models, while the proposed method, results in smaller systematic uncertainties. The DANSS result on this ratio
is one of the most precise among reactor neutrino experiments, yet does not significantly contradict to any previous measurement.

\end{abstract}


\maketitle


\section{Introduction}

Starting from the original idea of the reactor antineutrino anomaly (RAA, \cite{Mention:2011rk}) almost all reactor neutrino experiments make attempts to compare their results
to the predictions of various models, describing antineutrino production in reactor cores. Such models for the power reactors typically operate with the antineutrino
fluxes and spectra, produced in the decay chains from the fission of the four major isotopes, taking part in the reactor operation. Namely, these isotopes are: 
$^{235}$U, $^{239}$Pu, $^{238}$U and $^{241}$Pu, and corresponding values in this paper will have indices of 5, 9, 8 and 1, which is a common notation in many works.

As time goes on, the experiments accumulate more neutrino events and reduce their measurement uncertainties \cite{DayaBay:2022orm,DayaBay:2025fuel,RENO:2020dxd,Alekseev:2024mkb},
while the models improve their experimental base and analysis approaches \cite{Estienne:2019ujo,Perisse:2023efm,Kopeikin:2021ugh}. In spite of this tremendous progress 
the discrepancies in the absolute yield  and energy spectrum shape between models and experiments still persist. Also it is worth mentioning that the original conversion 
model by Huber and Mueller \cite{Huber, Mueller}, though did not undergo much changes, yet remains the most often referenced source for comparisons.

An important characterization of reactor antineutrino models can be made in terms of the inverse beta decay (IBD) yields, i.e., the number of antineutrinos per fission 
of a certain isotope obtained as the integral of the antineutrino spectrum from the fission products of this isotope multiplied by the IBD cross-section. 
Following the notation by the Daya Bay (DB) collaboration, we designate these values as $\sigma_i$, 
where the index $i$ corresponds to one of the four major isotopes mentioned above. The absolute values of these yields can be readily derived from the models, but their
estimates from the experimental data would rely on the understanding of the antineutrino detection efficiency, and thus may contain significant uncertainties.

The ratio of the yields for the two most significant isotopes, $\sigma_5/\sigma_9$, can also be very important in understanding of the role of each isotope in the nature
of the discrepancies between models and experiments. Though in an experimental determination of such ratio the detection efficiency should mainly cancel out, it arises
as a correlated uncertainty and has to be accurately accounted for.

In this paper we propose a novel simple way to directly determine the $\sigma_5/\sigma_9$ IBD yield ratio based on the analysis of the detector counting rate evolution 
with the fuel burnup throughout the reactor campaign. In this method the detection efficiency is naturally excluded from the consideration and thus the resulting uncertainty 
is lower, than that described above. We also present the numerical estimate of $\sigma_5/\sigma_9$ based on more than 7 years of data taking by the DANSS detector 
\cite{Alekseev:2024mkb}, including 5~almost full reactor campaigns. The result appears to have the smallest uncertainty compared to that from the other experiments, 
detecting antineutrinos from the power reactors. In this data analysis we follow the approaches by DB~\cite{DayaBay:2017jkb,DayaBay:2022jnn}
in order to preserve a direct comparability of the results.

\section{Method}

The detector count per fission is proportional to the linear combination of the individual isotopic yields:
\begin{equation}
N = \alpha \cdot \sigma_f = \alpha \cdot \left ( \sigma_8 f_8 + \sigma_1 f_1 +\sigma_5 f_5 + \sigma_9 f_9 \right ) ,
\label{form:1}
\end{equation}
\noindent
where $f_i$ are the corresponding fission fractions with their sum normalized to unity. The proportionality coefficient $\alpha$ includes the reactor and the detector geometry,
the number of protons in the detector, the detection efficiency and other components, independent of the fission fraction changes throughout the campaign.

Next, consider the derivative of $N$ on the $^{239}$Pu fission fraction $f_9$:
\begin{equation}
\frac{dN}{df_9} = \alpha \cdot \frac{d\sigma_f}{df_9} = \alpha \cdot \left ( \sigma_8 \frac{df_8}{df_9} + \sigma_1 \frac{df_1}{df_9} +\sigma_5 \frac{df_5}{df_9} + \sigma_9 \right ).
\label{form:2}
\end{equation}
\noindent
Here $\frac{d\sigma_f}{df_9}$ is the measure of the change in the total IBD event yield per unit of $^{239}$Pu fission fraction, and this is exactly the slope, discussed in 
the DB papers (see formula (4) in \cite{DayaBay:2017jkb}, for instance). All the derivatives in (\ref{form:2}) are considered as averages for full fuel campaigns, see details 
in the following section.

Now, divide (\ref{form:2}) by (\ref{form:1}), and then, in the right part, divide both the numerator and the denominator by $\sigma_9$: 
\begin{equation}
\frac{\frac{dN}{df_9}}{N} = \frac{\frac{d\sigma_f}{df_9}}{\sigma_f} = \frac{ \frac{\sigma_8}{\sigma_9} \frac{df_8}{df_9} + \frac{\sigma_1}{\sigma_9} \frac{df_1}{df_9}
 + \frac{\sigma_5}{\sigma_9} \frac{df_5}{df_9} + 1 }{ \frac{\sigma_8}{\sigma_9} f_8 + \frac{\sigma_1}{\sigma_9} f_1 + \frac{\sigma_5}{\sigma_9} f_5 + f_9}.
\label{form:3}
\end{equation}
\noindent
In the following discussion we will designate ${\frac{d\sigma_f}{df_9}}/{\sigma_f}=S_n$ and call it the ‘normalized evolution slope’, following \cite{DayaBay:2022jnn}.
Similar quantities are used in \cite{DayaBay:2017jkb} for partial slopes in smaller energy bins.

It only remains to express the ratio of interest from~(\ref{form:3}):
\begin{equation}
\frac{\sigma_5}{\sigma_9} = -\frac{\frac{\sigma_8}{\sigma_9} \left ( S_n f_8 - \frac{df_8}{df_9} \right ) + \frac{\sigma_1}{\sigma_9} \left ( S_n f_1 - \frac{df_1}{df_9} \right )
+ \left ( S_n f_9 - 1 \right ) }{ S_n f_5 - \frac{df_5}{df_9}}.
\label{form:4}
\end{equation}

\section{Numeric estimates}

The numerical parameters in the expression~(\ref{form:4}) can be split into several groups. Table~\ref{tab:1} presents the values of the parameters 
with their uncertainties and corresponding contributions to the final result error.

\begin{table*}
\caption{\label{tab:1} Numeric parameters for the $\sigma_5/\sigma_9$  calculation.}
\begin{tabular}{llll}
\hline
Parameter & Value & Source & Contribution to $\sigma_5/\sigma_9$ uncertainty\\
\hline
$\sigma_8$ & 10.1$\pm$1.0  & HM DB \cite{Huber,Mueller,DayaBay:2017jkb,Hayes:2017res} &  \\
$\sigma_1$ & 6.04$\pm$0.6  & HM DB \cite{Huber,Mueller,DayaBay:2017jkb,Hayes:2017res} &  \\
$\sigma_9$ & 4.36$\pm$0.11 & HM DB \cite{Huber,Mueller,DayaBay:2017jkb,Hayes:2017res} &  \\
\hline
$\sigma_8/\sigma_9$ & 2.32$\pm$0.24 & HM DB \cite{Huber,Mueller,DayaBay:2017jkb,Hayes:2017res} & 0.014 \\
$\sigma_1/\sigma_9$ & 1.39$\pm$0.14 & HM DB \cite{Huber,Mueller,DayaBay:2017jkb,Hayes:2017res} & 0.039 \\
\hline
$df_8/df_9$ & 0.0369 & KNPP & \multirow{6}*{
\hspace{-1.1em}$\left.\begin{array}{l} \\ \\ \\ \\ \\ \\ \end{array}\right\}$
\begin{minipage}[c]{4cm}
\begin{flushleft} 
0.0003 (Campaign spread) \\ 
0.0006 ($f_5\cdot1.05$) \\ 
$-0.0016\;$($f_5\cdot1.00\to f_5\cdot1.05$) \\ $-0.0020\;$(Nonlinearity)
\end{flushleft}
\end{minipage}
} \\
$df_1/df_9$ & 0.2803 & KNPP & \\
$df_5/df_9$ & $-1.3171$ & KNPP & \\
$f_8$ & 0.0718 & KNPP & \\
$f_1$ & 0.0555 & KNPP & \\
$f_5$ & 0.5727 & KNPP & \\
\hline
$S_n$ & $-0.3799\pm$0.0317\hspace{1em} & DANSS \cite{Alekseev:2024mkb} & 0.041 \\
\hline
\boldmath${\sigma_5/\sigma_9}$ & \textbf{1.529} & \textbf{Total uncertainty:} & \textbf{0.057} \\
\hline
\end{tabular}
\end{table*}

\subsection{Minor IBD yield ratios}
The two minor fission fractions of $^{238}$U and $^{241}$Pu yet give a notable contribution to the numerator of (\ref{form:4})
and have to be taken into account. For the full compatibility with the DB results the numerical values of $\sigma_8/\sigma_9$ and $\sigma_1/\sigma_9$ 
with the corresponding uncertainties are taken as the HM model \cite{Huber, Mueller} estimates from~\cite{DayaBay:2017jkb}. A nice summary of the DB assumptions 
in~\cite{DayaBay:2017jkb} is presented in Table~II of~\cite{Hayes:2017res}. The uncertainties in the mentioned yield ratios are totally dominated 
by those of $\sigma_8$ and $\sigma_1$, and thus treated as uncorrelated in this paper.

\subsection{Fission fractions and derivatives}
The fission fractions were provided for the DANSS experiment by the personnel of the Kalinin nuclear power plant (KNPP) for almost 5 full fuel campaigns, 
see Figure~\ref{fig:0} for the plots.
The data on the fuel evolution and the corresponding fission fraction values are based on the validated evolution code \mbox{BIPR-7A}, which is a part of
the CASCADE software package~\cite{cascade}. \mbox{BIPR-7A} is designed for fast operational calculations and simulates the reactor power history, starting 
from the loading map with fresh fuel and partly burned cassettes at the beginning of each campaign. The power distributions are verified against the readings
of the direct charge detectors and the typical deviations at the cassette level are in the range of (0.7--2.8)\% r.m.s. \cite{Zhutikov:2024fhi, KNPP_priv}.
The discrepancies in the power calculations do not directly cast the limits on the uncertainties of the fission fractions. Alternatively the estimates
are obtained from the comparison of the isotopic content from the destructive analysis of the spent fuel to that from several simulations. The discussion
of this approach can be found in~\cite{DayaBay:2017jkb,Barresi:2023sfw} (and references therein), which states $\Delta f/f$=5\% as a conservative estimate 
of the major fission fraction uncertainty. Similar studies were performed with the spent cassettes from Russian WWER reactors and similar 5\% unceratainties
were reported from these measurements~\cite{KNPP_priv}.

The values of the fission fraction parameters in Table~\ref{tab:1} are their averages over the five fuel campaigns. The derivatives are calculated by linear 
fits of the corresponding dependencies for each individual campaign and the slope of the fit is taken as the campaign average. Following the approach by DB,
all the other fission fractions are taken at the value of $f_9$ equals exactly 0.3. The contribution from these parameters to the final uncertainty is 
estimated by four ways:

\begin{figure}
\includegraphics[width=\columnwidth]{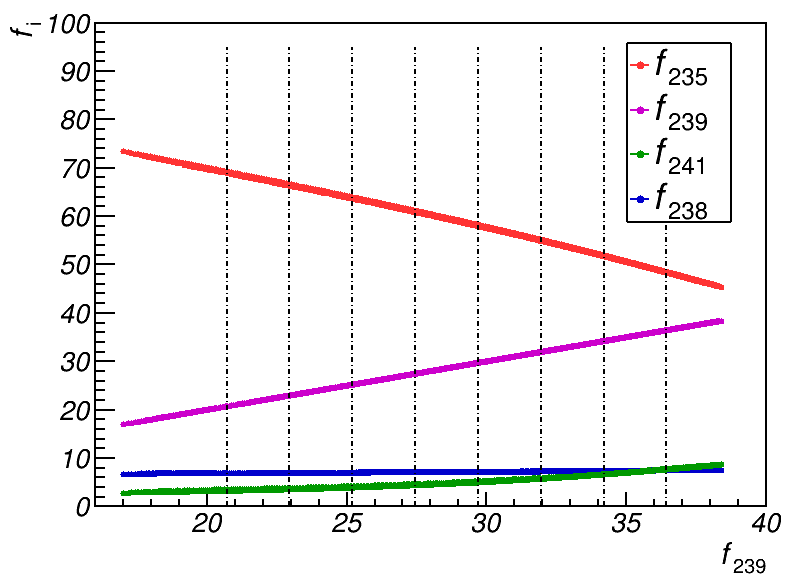}
\caption{\label{fig:0} Fission fractions of the four major isotopes, superimposed for all campaigns; the spread is within the line thickness. 
Vertical lines separate the $f_9$ intervals for the data averaging.}
\end{figure}                                                                                                                                                                                                                                                                                                                                                                                                                                                                                                        

    \begin{itemize}
    \item the spread of the resulting $\sigma_5/\sigma_9$ values for 5 campaigns, when all parameters for a certain campaign are substituted into (\ref{form:4});
    \item the change of the result when the $f_5$ fission fraction is increased by $\Delta f_5/f_5$=5\% in the course of all the 5 fuel campaigns. The other major
fraction $f_9$ is changed accordingly to keep the normalization of the fractions sum to unity. The minor fractions are left intact;
    \item a similar procedure, but with no changes in $f_5$ at the beginning of each campaign which grow to 5\% at the campaign ends;
    \item the linear fits of the fission fraction dependencies are replaced by quadratic ones and the derivatives are taken as the tangent slopes at $f_9=0.3$.
    \end{itemize}
All these contributions appear to be negligibly small and do not notably add to the final uncertainty, even if all four are summed in quadratures. 

\subsection{Normalized slope}
The dependence of the IBD count rate on the $^{239}$Pu fission fraction was derived on the basis of the DANSS detector 
antineutrino sample, accumulated during 7.5~years of operation. By this point DANSS recorded and analyzed about 8~million antineutrino events \cite{Alekseev:2024mkb},
which makes it the largest antineutrino sample among all reactor experiments, while the signal to correlated background ratio exceeds the value of 50. The procedure
of the event selection and of the background subtraction is described in~\cite{DANSS:PLB18}. The most significant backgrounds -- the accidental background (ACB) and
the cosmic background (CB) are calculated from data for each measurement point of 2-3~day duration, thus reflecting changes due to the detector position
and to the reactor operations. The ACB comprises approximately 20\% of the IBD rate, while its subtraction is a statistically sufficient and mathematically strict procedure,
which naturally follows the changes in the ACB amount and spectra, no matter how large they are. The muon flux variations do not exceed 6\%. Most of the cosmic 
background events are rejected by the active veto system, while what escaped the veto tagging was proved to have the same spectrum. The fraction of such 
events is calculated using the reactor-off spectra, while the actual amount of the CB is obtained by multiplying of the current veto-tagged spectrum by 
this fraction. The drift of the veto efficiency and its variations due to the muon angular distribution changes at different detector positions are 
additionally accounted through the 'veto transparency' parameter, also calculated from the actual data. The cosmic background is about 1.8\% of the 
IBD rate at the top detector position and its spectrum is well determined statistically.
The energy spectrum of the background from the fast neutrons produced outside the detector shielding is estimated by a linear extrapolation from the (10--16)~MeV region to
lower energies. This background constitutes $\sim$0.1\% of the IBD signal. The contribution from $^8$Li/$^8$He induced by cosmic muons is also evaluated below 0.1\%, while
this background component is currently not subtracted.

The primary physics goal of the DANSS experiment is the search for sterile neutrinos, and this requires taking data at various distances from the neutrino source.
The IBD counts used for the slope analysis were all normalized to the position of the detector, closest to the reactor core, using a toy geometrical MC reflecting
the spread of the neutrino production and detection points. The detector position is controlled at the level of few millimeters with the help of the laser 
pointers located on the four corners of the detector movable frame and the rulers, attached to the fixed columns of the lifting mechanism. The movement of 
the core burning center does not exceed several centimeters at full reactor power, but shows a correlated behavior with the campaign time. The burning
center changes are taken into account based on the 'offset' parameter values, provided by the KNPP staff. Offset is defined as the power difference between 
the bottom and top halves of the reactor core divided by the total power.
Only the data at full reactor power were taken into the slope analysis in order to exclude significant movements of the burning center due to the control rod insertions.
The influence of the adjacent reactors does not exceed 0.6\% and is taken into account by the corresponding change in the positron spectrum normalization.

The DANSS detector~\cite{DANSS:design} is composed of plastic scintillator strips which show excellent stability of the light yield with only 0.55\% decrease per 
year~\cite{DANSS:ageing}. The gadolinium for the neutron capture is bound in the reflective layer of the strips which is co-extruded with the scintillating body,
and thus the gadolinium concentration is fixed by design. The energy response only drifts due to the temperature variations and this is being 
continuously calibrated~\cite{DANSS:Calibration}: fast correction of the SiPM response is based on their noise spectra and is updated every half an hour, while the ‘absolute’ 
calibration of both SiPMs and PMTs is made using the vertical muons and this happens every two days. Naturally, the calibration procedure accounts 
for the small light yield deterioration. The actual maps of the dead channels are built from the data and reflect the status of the channel failures and repairs.
The maps are used for the MC simulations of the IBD events, which then undergo the full analysis procedure equivalent to that for the experimental data.
The variations of the detection efficiency due to the dead channels are taken into account. Numerically they are below 3\% and show no correlations with the campaign time. 
The dead time of the detector arises at the analysis stage due to the trigger isolation and muon veto cuts. The dead time is proportional to the raw trigger rate and depends 
on many external factors, however it is accurately accounted for by the analysis software. The live time variations are in the range (84--88)\%. The IBD rate changes
in the DANSS detector are illustrated by Figure~\ref{fig:A1} in the Appendix, all corrections made, all backgrounds subtracted. 

To keep the similarity with DB, the ‘per fission’ normalization of the counts was performed 
using the same values of fission energies~\cite{Ma:2012bm}. Unlike DB, DANSS detects neutrinos from a single power reactor, since it is located at a short 
distance of 10.9 to 12.9 meters from its core center. This means, that each single IBD rate measurement, which in case of DANSS corresponds to an approximately
3~day period, and has a statistical accuracy of about 1\%, can specifically be attributed to a certain $^{239}$Pu fission fraction. Yet to maintain the full compatibility
with the DB approach \cite{DayaBay:2017jkb,DayaBay:2022jnn}, all measurements were combined into 9 groups with certain $f_9$ intervals, see Figure~\ref{fig:0}. 
The range of the $f_9$ values is almost twice wider in the DANSS experiment compared to that in the DB analysis. 

The estimate of the systematic uncertainty of the IBD rate measurement by DANSS is based on the long term observation of the reactor power. After the subtraction
of all backgrounds and applying all corrections as described above, the detector counts are additionally adjusted to accommodate for the changes of the isotopic fuel content
using the H-M model~\cite{Huber, Mueller}, and then compared to the reactor power data, obtained from the conventional NPP instrumentation. The correspondence of the IBD counts 
to the reactor power was once established based on one month measurement in October 2016, and then this single normalization was used for all 7.5 years of the observation period.
The analysis of the deviation between the two power measurement methods~\cite{DANSS:FissFract} estimated the combined systematic uncertainty as 0.8\%. This number
includes both the remaining uncontrollable systematics from DANSS and the uncertainties of the power measurements by the NPP instrumentation. The latter alone 
is evaluated by the NPP staff to have the relative accuracy not better than 1\%, while the observations by DANSS insist on a better estimate.
For the slope measurements, we have to normalize the DANSS counts by the reactor power, thus adding the uncertainties of the NPP power measurements to the intrinsic DANSS
systematics. Thus the total systematic uncertainty of the IBD rate result exactly equals to the quoted combined value of 0.8\%.

This number should be treated as a systematic uncertainty of a single 2-3 day rate measurement, and is supposed to average down to a smaller value when such measurements
are combined into $f_9$ groups, according to the number of combined points. Yet this is only true if all the averaged results are fully uncorrelated, which is not
the case of DANSS. The uncertainty under discussion contains an unknown but, presumably, large contribution from the NPP power measurements, which has a specific
behavior. The resulting power value is a weighted average from four completely independent subsystems, using both the heat flow balance and the neutron field
intensities. A complicated operational procedure is used to estimate and monitor the reliability and the actual measurement uncertainty of each subsystem. Based 
on these estimates, a decision is made from time to time to change the weights, or even temporarily exclude a subsystem from averaging. The procedure also includes
cross-calibrations of the subsystems. Such changes can be observed as small jumps in the IBD rates after periods of stability or small slow drift, and the duration
of such periods is typically 1-3 months. In terms of the DANSS rate analysis we will speak of the uncertainty {\em correlation latency} of the order of months,
meaning that the consecutive 2-3 day rate measurement points may have partial common correlated uncertainty, while we do not expect correlations between the groups
of data from different campaigns, which are separated by almost 1.5~years.

While the systematics of the rate measurements is well understood, it is very difficult to quantify the contribution of the NPP power uncertainties, as well
as their correlation latency. Instead, it looks more natural in this case to extract the uncertainty estimate from the data itself assuming that the results 
from different campaigns are fully uncorrelated. In this analysis the combining of the data points into $f_9$ groups was made separately for each fuel campaign and 
the r.m.s. spread of the results between the campaigns was added in quadratures to the corresponding statistical errors. 

The resulting campaign-averaged points are shown in Figure~\ref{fig:1} with full circles together with their linear fit (solid line), while Figure~\ref{fig:A2} 
in the Appendix gives an impression of similar results for the four full individual campaigns. The IBD rates are normalized to their average value taken as the linear 
fit result at $f_9=0.3$. The $\chi^2$ of the fit is 5.9 per 7~degrees of freedom, and this can be compared to the value $\chi^2=70.2$ that results from the fit with pure 
statistical errors, see Figure~\ref{fig:A3}a. Obviously, the uncertainty of each $f_9$ point, and, consequently, of the normalized slope, is dominated 
by the systematic errors. Another rough estimate can be made assuming that the quoted 0.8\% uncertainty averages out according to the number of uncorrelated campaigns.
Adding of 0.8\%/$\sqrt{5}$ to the statistical errors results in the fit $\chi^2=6.8$ (Figure~\ref{fig:A3}b), which is in good agreement with that 
with the uncertainties from the campaign spread and proofs the similarity of the two approaches. For comparison and as a reference, 
all figures also show the DB data points (empty circles), taken from~\cite{DayaBay:2017jkb}, but converted to the normalized form, along with the 
linear fit of these points performed in this work. The slopes in Figure~\ref{fig:1}, though similar, differ beyond $1\sigma$ consistency, indicating a higher
dependence on the fuel evolution in the case of the DANSS results (see below for numbers).

\begin{figure}
\includegraphics[width=\columnwidth]{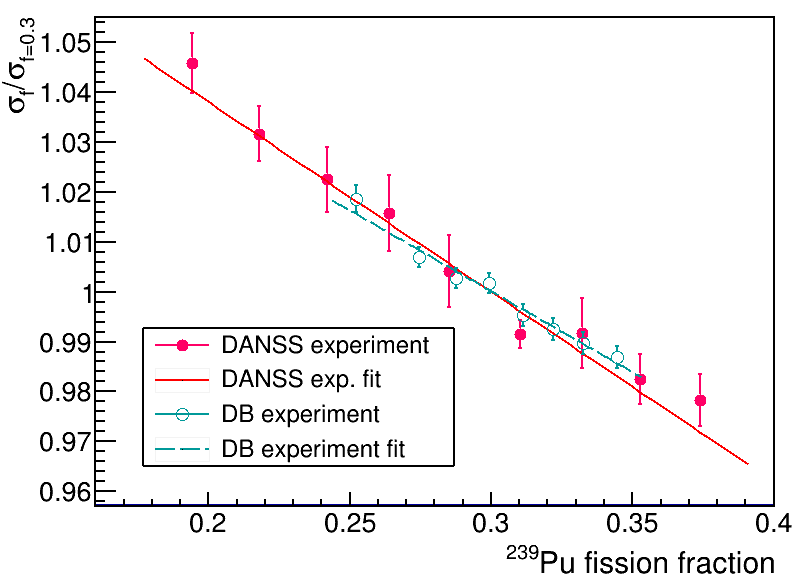}
\caption{\label{fig:1} Count rate of the DANSS detector as a function of the $^{239}$Pu fission fraction (full circles) and it’s linear fit (solid line);
similar measurements from the DB experiment~\cite{DayaBay:2017jkb} (empty circles and dashed line).  All data are normalized by corresponding linear fit values 
at $f_9$=0.3.}
\end{figure}                                                                                                                                                                                                                                                                                                                                                                                                                                                                                                        

As it follows from the Table~\ref{tab:1}, the experimental measurement of the normalized slope $S_n$ gives the main contribution to the total resulting 
uncertainty in the $\sigma_5/\sigma_9$ ratio, while it will be limited by the $\sigma_1/\sigma_9$ estimate if the slope measurements are improved.

\subsection{$\sigma_5/\sigma_9$ yield ratio}
Table~\ref{tab:2} compares values of $\sigma_5/\sigma_9$ ratio, obtained from various sources by means of several methods. The first four values of the ratio
are obtained using formula~(\ref{form:4}) from the normalized slope values $S_n$ given in the left column. The first line once again shows the result from 
DANSS~\cite{Alekseev:2024mkb} for better comparison. The next line is based on our fit of the Daya Bay data from~\cite{DayaBay:2017jkb}, scaled to 
the normalized form, see Figure~\ref{fig:1}. In the third line the normalized slope value is obtained by the direct division of the absolute slope 
$d\sigma_f/df_9$ by the absolute average yield $\sigma_f$  from that same paper~\cite{DayaBay:2017jkb}, assuming uncorrelated errors of these two quantities.
The latter presumably leads to an overestimate of the error, since both values may contain, for example, the common uncertainty in the detection efficiency.
The estimate on $S_n$ in the fourth line is obtained from the most recent DB numbers 
in~\cite{DayaBay:2025fuel} in a similar way and produces the value of the yield ratio in good agreement with the DANSS result. The $\sigma_5/\sigma_9$ ratio 
in line five is obtained by the direct division of the absolute isotopic yields from~\cite{DayaBay:2017jkb}. This recipe is used in~\cite{Hayes:2017res}, 
again with the assumption of the uncorrelated errors. The above notice about common uncertainties remains true here either. The ratio value in the next line is 
taken directly from~\cite{DayaBay:2025fuel} as provided by DB and shows almost an exact coincidence with our number in line four based on the DB result on $S_n$.
For completeness, the following line shows the result by KI group~\cite{Kopeikin:2021ugh}. The last line shows the prediction of the Huber and Mueller 
model~\cite{Huber,Mueller} as interpreted by DB in~\cite{DayaBay:2017jkb} and summarized in~\cite{Hayes:2017res}.

\begin{table}
\caption{\label{tab:2} Comparison of the normalized slope $S_n$ from several sources and several $\sigma_5/\sigma_9$  estimates.}
\begin{tabular}{p{0.26\columnwidth}p{0.45\columnwidth}p{0.22\columnwidth}}
\hline
Normalized slope $S_n$ & Source & \mbox{IBD yield} \mbox{ratio} $\sigma_5/\sigma_9$ \\
\hline
$-0.380\pm0.032$ & DANSS \cite{Alekseev:2024mkb} & 1.529$\pm$0.057 \\
$-0.324\pm0.029$ & DB \cite{DayaBay:2017jkb}, this work fit & 1.459$\pm$0.052 \\
$-0.315\pm0.031$ & DB \cite{DayaBay:2017jkb}, uncorrelated errors & 1.448$\pm$0.054 \\
$-0.336\pm0.023$ & DB \cite{DayaBay:2025fuel}, uncorrelated errors & 1.473$\pm$0.048 \\
\hline
& DB \cite{DayaBay:2017jkb,Hayes:2017res} & 1.445$\pm$0.097 \\
& DB \cite{DayaBay:2025fuel} & 1.48$\pm$0.07 \\
& KI \cite{Kopeikin:2021ugh} & 1.45$\pm$0.03 \\
& HM DB \cite{Huber,Mueller,DayaBay:2017jkb,Hayes:2017res} & 1.53$\pm$0.05 \\
\hline
\end{tabular}
\end{table}

All Daya Bay results based on the normalized slope are close to each other and, what is most interesting, almost coincide with their estimates from the isotopic
yields. Mind that the fuel parameters in formula~(\ref{form:4}) are taken for a single WWER-1000 reactor at KNPP, while used for the data averaged over 
6~pressurized water reactors at Daya Bay, though having similar power, but of a completely different model. This is probably not a simple coincidence, but
rather a manifestation of a certain universality of the proposed approach for all power reactors of this type. We already showed that the current DANSS 
result only very slightly depends on the variations of the fission fraction data, be it the campaign spread or intentional variations within the estimated
uncertainties. Moreover, power reactors of various models have similar values of the fission fractions in the middle of the fuel campaign, where these
values are taken in formula~(\ref{form:4}). The derivatives over $f_9$  should also have similar values because of the similar nature of $^{239}$Pu
production throughout the $^{235}$U burnup.

\begin{figure}
\includegraphics[width=0.4\textwidth]{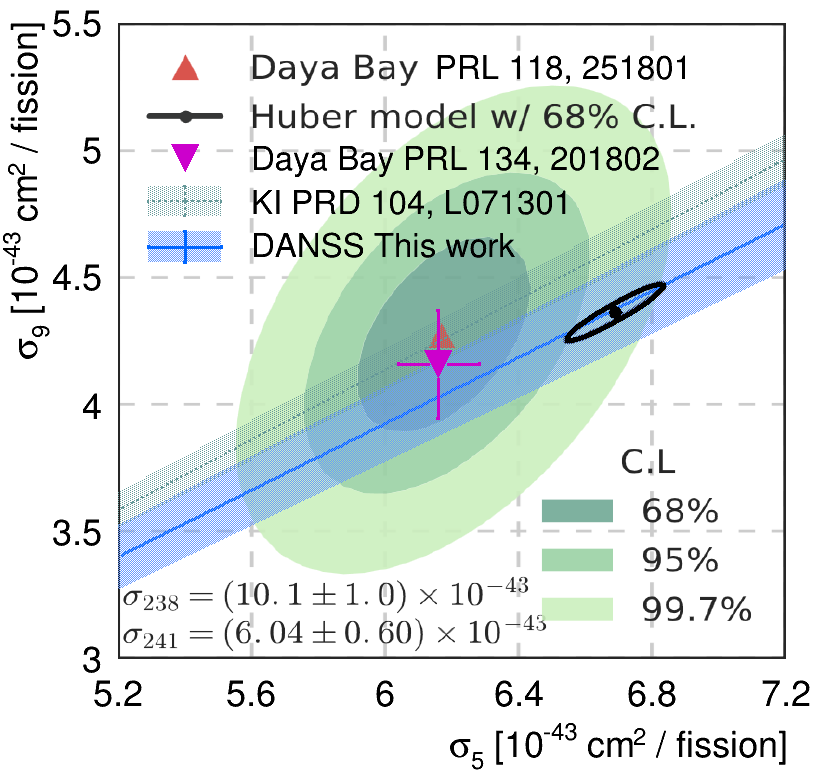}
\caption{\label{fig:2} Daya Bay results on the $^{235}$U and $^{239}$Pu IBD yields from the original~ \cite{DayaBay:2017jkb} (green ellipses) and
final~\cite{DayaBay:2025fuel} (pink triangle with cross) analyses; black dot and ellipse give the H{-}M model 
predictions~\cite{Huber,Mueller,DayaBay:2017jkb,Hayes:2017res}; lines with $1\sigma$ error corridors represent the $\sigma_5/\sigma_9$ ratio obtained by 
KI~\cite{Kopeikin:2021ugh} (dotted cyan) and the result of this work (solid blue)}.
\end{figure}

At the same time the DANSS result, though does not contradict to those from DB, yet obviously prefers the value, predicted by the H-M model. 
This is illustrated in Figure~\ref{fig:2}, where the blue band showing $\sigma_5/\sigma_9$ from this paper is superimposed on top of the original plot
from~\cite{DayaBay:2017jkb} presenting the individual isotopic IBD yields (red triangle). The figure gives an idea of the error correlations by means
of the confidence level ellipses (green). The H-M model prediction is also given in the original plot (black), while the most recent DB 
values~\cite{DayaBay:2025fuel} are superimposed as a pink reversed triangle with a cross. The estimate by the KI 
group~\cite{Kopeikin:2021ugh} is shown with the dotted line and surrounding error corridor. It is only in $1.2\sigma$ disagreement with the DANSS data.

It is worth mentioning that the errors in the $\sigma_5/\sigma_9$ quantity are 1.5-2 times smaller in case of its determination from the normalized slope
compared to the direct IBD yield ratio. Partly this is because in the direct approach the individual errors of the isotopic yields are treated
as uncorrelated, while this is obviously not true -- both yields, at least, contain common uncertainty in the detection efficiency. A correct account for the
error correlations may decrease the error estimate in this case. At the same time the method proposed in this paper is free from many of these problems
from the very beginning as it operates with relative quantities.

\section{Conclusions}

We presented a method to directly determine the isotopic inverse beta decay yield ratio $\sigma_5/\sigma_9$ based on the analysis of the relative changes in the detector
counting rate throughout the power reactor fuel campaign. The procedure features smaller estimated errors in spite of the fact that many parameters from several
sources are used for the calculations. Numerical estimates based on about 8 million antineutrinos recorded in more than 7 years of DANSS detector operation
indicate slightly higher dependence of the counting rate on the reactor fuel evolution than measured by the Daya Bay experiment, the normalized slope is 
in (1.1-1.4)$\sigma$ discrepancy. The resulting value of the $\sigma_5/\sigma_9$ ratio shows the smallest uncertainty
among the other neutrino experiments at power reactors, coincides with the predictions of the H-M model, agrees with the DB numbers, 
while the discrepancy with the KI results is about 1.2 standard deviations.  Further improvements 
in the normalized slope accuracy, if used in the framework of the suggested approach, will also require better understanding of the isotopic yields 
of the minor fission fractions, $^{241}$Pu in the first place. Nevertheless, the proposed method gives an independent way to determine 
the $\sigma_5/\sigma_9$ IBD yield ratio and can be applied to the data from the large variety of commercial power reactors used for neutrino experiments. 

\begin{acknowledgments}

The DANSS collaboration deeply values the permanent assistance and help provided by the administration and staff of KNPP.
This work is supported in the framework of the State project Science by the Ministry of Science and Higher Education of the Russian Federation, 
Grant No. 075-15-2024-541.

\end{acknowledgments}

\vfill

\bibliography{DANSS_references}

\appendix*
\section{Additional figures}

\begin{figure*}
\includegraphics[width=\textwidth]{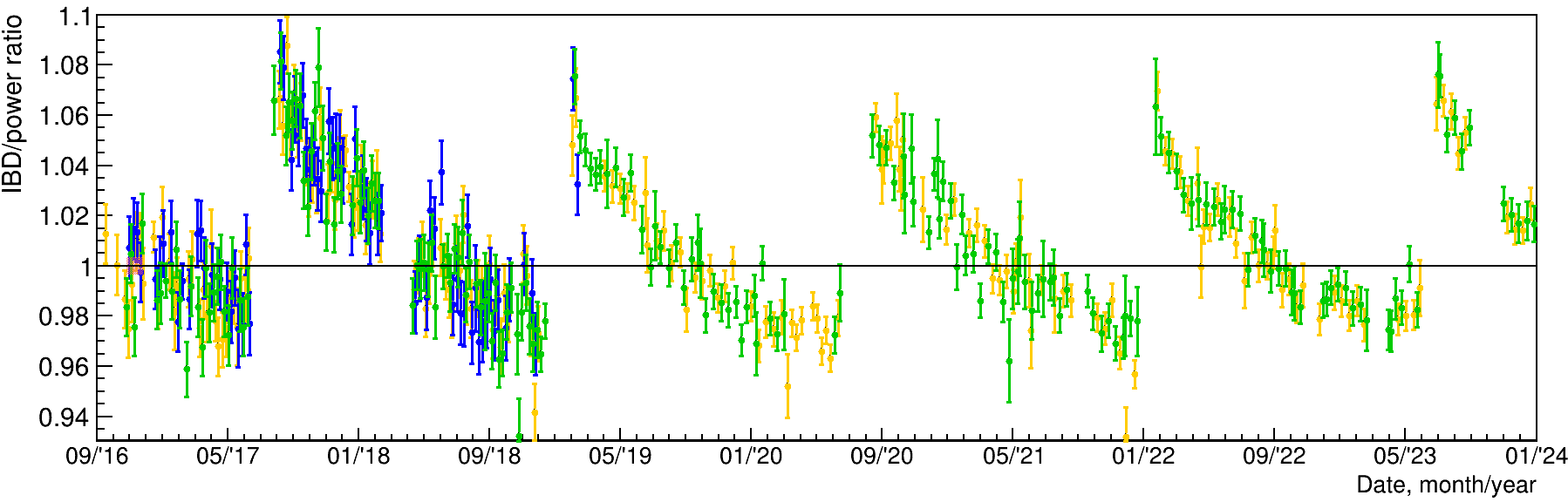}
\caption{\label{fig:A1} Relative IBD count rate of the DANSS detector, normalized by the reactor power, as a function of the calendar time, all backgrounds subtracted, 
all corrections applied, statistical errors only; orange, blue and green points correspond to the top (closest to the reactor), middle and bottom positions of the detector; 
only measurements at full reactor power presented. The position matching uses a toy geometrical MC of the reactor core and the detector. A one month period from November~18, 
2016 (shown with red dotted rectangle) is used to establish the correspondence between the IBD rate and the reactor power and this common normalization is used for the
whole analysis period. Four full fuel campaigns are clearly seen, separated by the reactor-off periods for the fuel reload; other blank gaps correspond to either pauses 
for the DANSS maintenance or the intervals of the reactor operation at low power.
}
\end{figure*}    

\begin{figure*}
\begin{tabular}{cp{1cm}c}
\includegraphics[width=0.85\columnwidth]{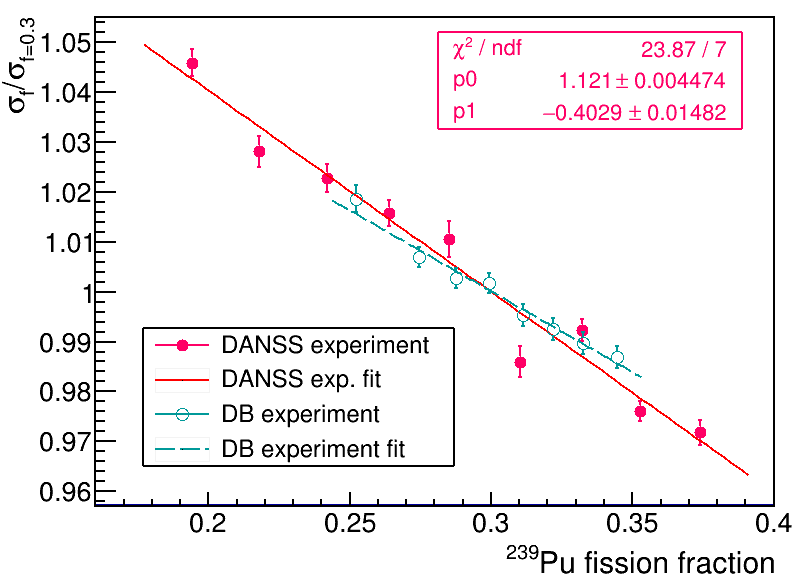} & &
\includegraphics[width=0.85\columnwidth]{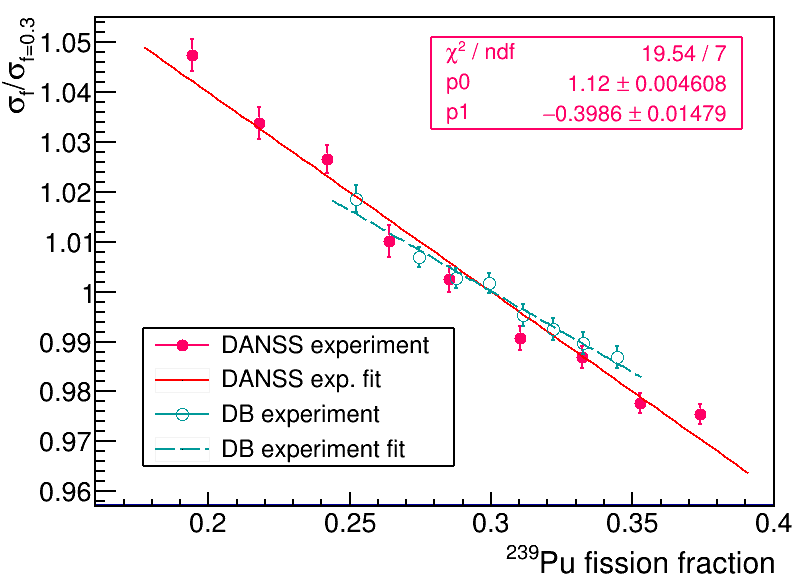} \\
a) & & b) \\
\includegraphics[width=0.85\columnwidth]{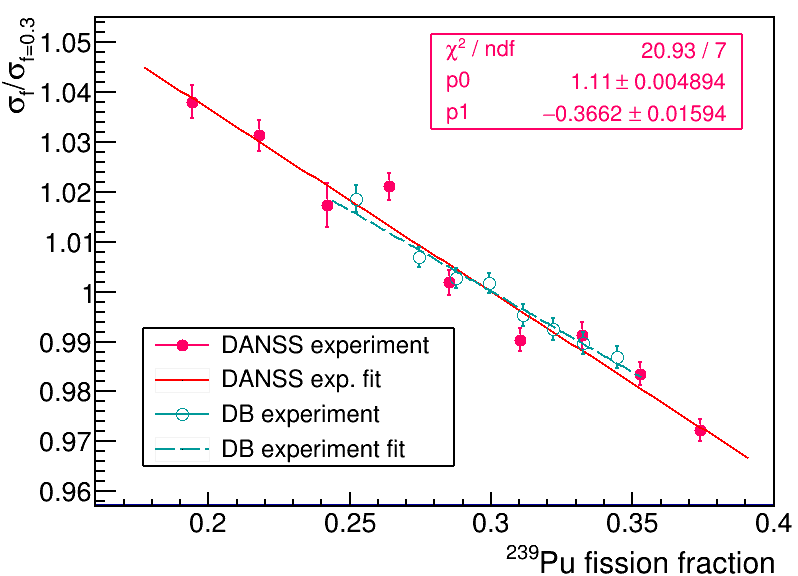} & &
\includegraphics[width=0.85\columnwidth]{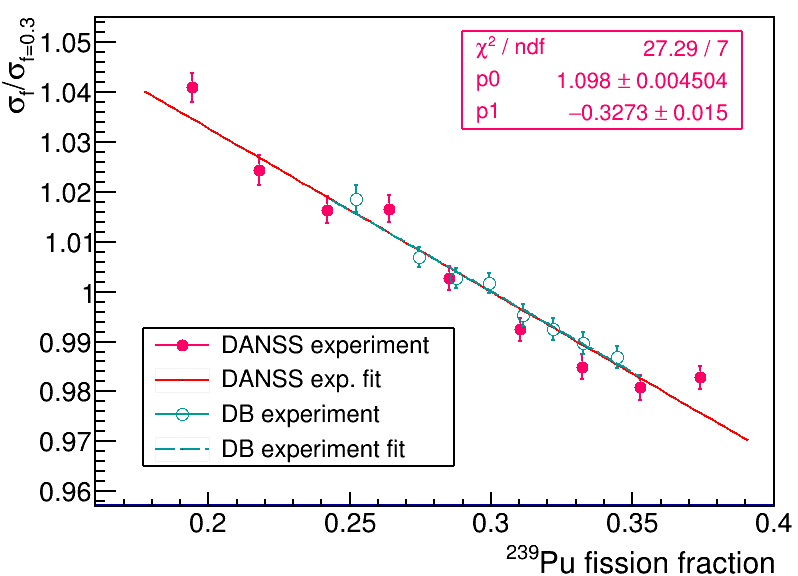} \\
c) & & d)
\end{tabular}
\caption{\label{fig:A2} Relative IBD count rates of the DANSS detector as functions of the $^{239}$Pu fission fraction (full circles) with their linear fits (solid lines) for
the four full fuel campaigns (a-d), statistical error only; empty circles and dashed lines repeat the DB result~\cite{DayaBay:2017jkb} for comparison.  All data are normalized 
by corresponding linear fit values at $f_9$=0.3.}
\end{figure*}    

\begin{figure*}
\begin{tabular}{cp{1cm}c}
\includegraphics[width=0.85\columnwidth]{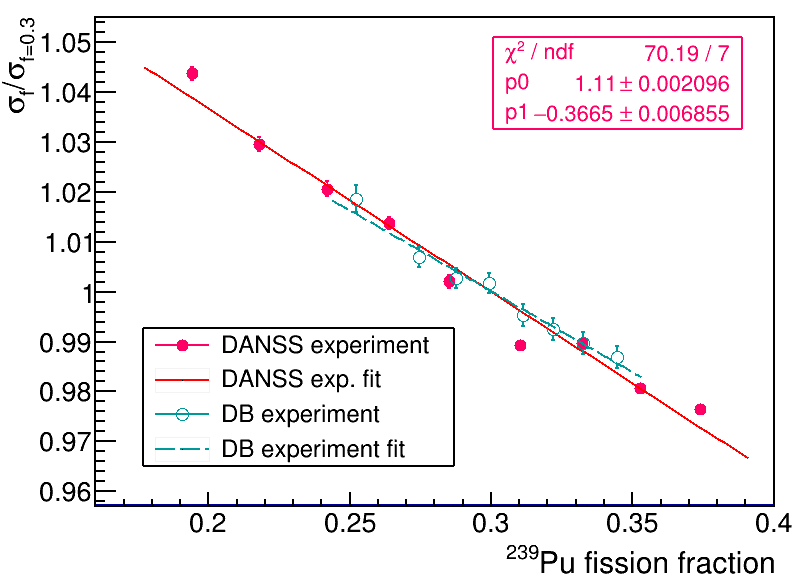} & &
\includegraphics[width=0.85\columnwidth]{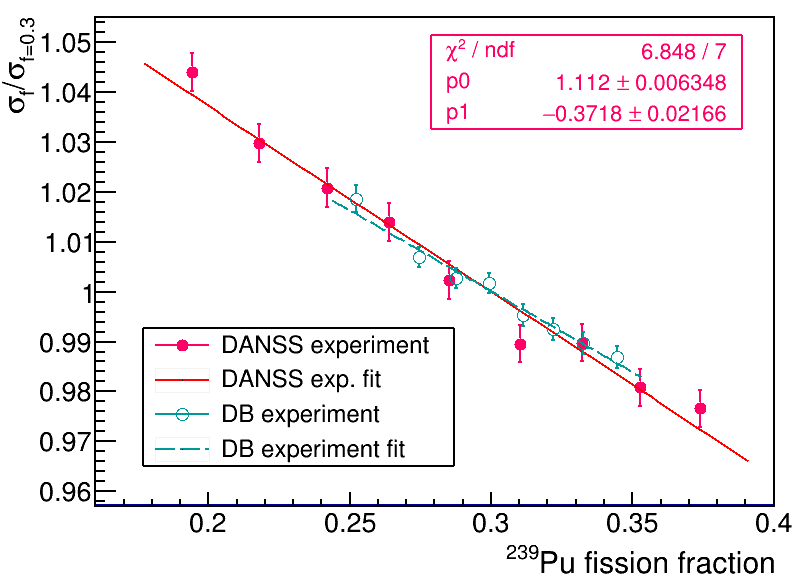} \\
a) & & b)
\end{tabular}
\caption{\label{fig:A3} Relative IBD count rates of the DANSS detector as functions of the $^{239}$Pu fission fraction (full circles) with their linear fits (solid lines) for the
total campaign averages with pure statistical errors (a) and with statistical errors increased by a factor of $0.8/\sqrt{5}$\% (b). Empty circles and dashed lines repeat 
the DB result~\cite{DayaBay:2017jkb} for comparison. All data are normalized by corresponding linear fit values at $f_9$=0.3.}
\end{figure*}

\end{document}